\documentclass[12pt,a4paper]{article}           
\title{\vspace{-40pt} \textbf{\large 
Crystallographic Method for Exact Describing Quasicrystal Structures\\
}}
\author{\vspace{-32pt}\\ \itshape V.C.~Gouliaev\footnote{Email: comp@akadem.ru} \\
\itshape\small L. V. Kirensky Institute of Physics, of Russian Academy of Sciences\\
\itshape\small Krasnoyarsk 660036, Russia}
\date{}
   
\usepackage{amssymb, amsmath}
\usepackage[dvips]{graphicx,color}
\usepackage{captnsty,multicol}

\renewcommand{\thesection}{\normalsize\arabic{section}}
\begin{document}
\maketitle
\renewcommand{\abstractname}{}
\begin{abstract}
\vspace{-24pt} 
In this paper the problem of the theory of a quasicrystal structures -
the determination of coordinates of each atom of quasicrystal
in analytical form - is solved. 
Within the framework of the proposed model a periodic crystal can be presented 
as a particular case of a quasicrystal.
The simple and explicit analytical formulas 
which describe the location of each atom in a quasicrystal are given.
The exact solutions for Penrose and Ammann-Beenker quasicrystal structures 
are given.  
On the basis of the analytical formulas the routines are created.
The routines are inserted directly into graphical files generating the 
quasiperiodic structures.
\end{abstract}
%
%
\section       {\normalsize Introduction}
\vspace{-6pt}
Since quasicrystals were discovered [1] quasiperiodic structures have been 
widely studied, see e.g. [3-14]. The two standard ways to describe these structures are the 
cut and project method [3-7], and the grid algorithm [8-10]. 
However the basic unsolved problem of the theories of quasicrystals was the 
problem: the determination of true atom locations in a quasicrystal [2].
The solution of this problem is given in this paper. To solve the specified 
problem a method has been applied, which 
includes traditional crystallography, which describes 
periodic and incommensurate crystalline structures. The symmetries of 
quasicrystal structures are considered with the superspace-groups [15, 16], 
which are the classical method of describing incommensurate structures now.
%
\section       {\normalsize Quasicrystal structures}
\vspace{-4pt}
Structures, in particular quasicrystals, which form point diffraction patterns, 
but aren't periodic, can be described by the finite number of structural
cells. The scheme for a two-dimensional 
quasicrystal with structural cells $\mathbf{a}$--$\mathbf{w}$ is given on fig.\,1.
\begin{figure}[h]
\centering             
\includegraphics*{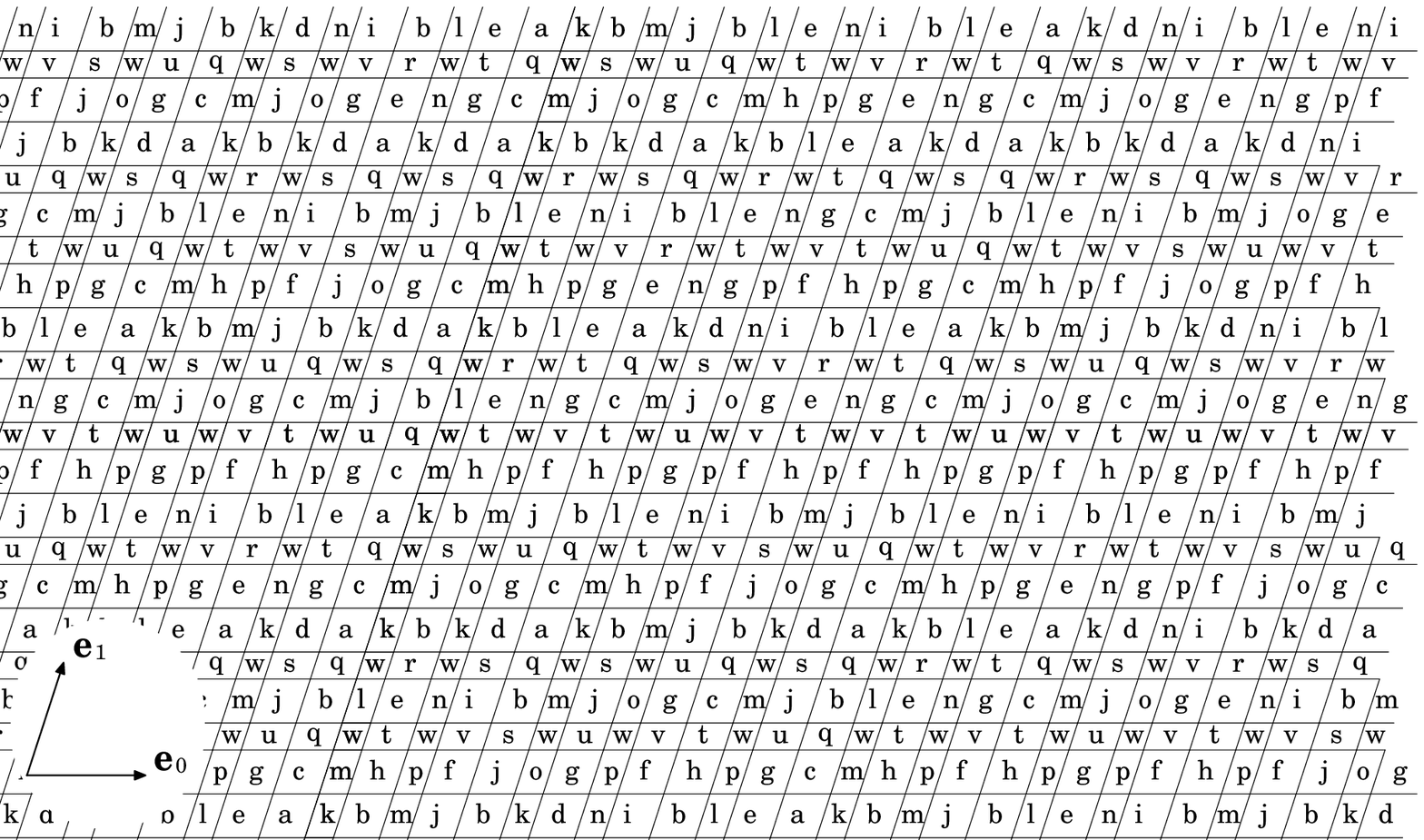}
\caption{The scheme of two-dimensional quasicrystal 
with structural cells $\mathbf{a}$--$\mathbf{w}$.}
\end{figure}
\\In such a quasicrystal with crystallographic axes, 
parallel $\mathbf{e}_0$, $\mathbf{e}_1$, structural cells are located
in the nodes of the quasiperiodic lattice with the coordinates
%
\begin{equation}
\mathbf{r_n}(\boldsymbol{\alpha,\beta})=
\mathbf{e}_0\,x_{n_0}\!(\alpha_0,\beta_0) 
+\mathbf{e}_1\,x_{n_1}\!(\alpha_1,\beta_1),
\end{equation}
%
where $x_{n_l}=a_l\left (n_l+\alpha_l + 
1/\chi_l \left\lfloor (n_l+\beta_l)/\sigma_l \right\rfloor\right )$, 
$\lfloor{x}\rfloor$ -- is the greatest integer part not exeeding $x$, 
$\sigma_l>1$ and $\chi_l$
-- irrational numbers, $\alpha_l$ and $\beta_l$ -- arbitrary numbers, 
$a_l$ -- the parameters of the lattice, $l=0,1$.
Further we will consider
\mbox{$\beta_l\left/\sigma_l\right.<1$}, since in the opposite case $\alpha$ 
changes, 
i.e. $\alpha_{l}'=\alpha_l+1/\chi_l{\left\lfloor{\beta_l/\sigma_l}\right\rfloor}$.
Introduce a vector 
$\boldsymbol{\Upsilon}_{\mathbf n}(\boldsymbol{\beta})$ in the Cartesian 
coordinates: 
\begin{equation}
\boldsymbol{\Upsilon}_{\mathbf n}(\boldsymbol{\beta}) = 
\left (\sigma_0\!\left\{\frac{n_0+\beta_0}{\sigma_0}\right\}, \;
\sigma_1\!\left\{ \frac{n_1+\beta_1}{\sigma_1}\right\}\right ),
\end{equation}
where $\{x\}$ -- the fractional part of $x$. Then the atom density distribution 
for quasicrystal with $K$ structural cells can be written in the form
\begin{align}
\label{math/3}
\rho(\mathbf{n},\boldsymbol{\alpha},\boldsymbol{\beta})=&
\sum_{n_0,\,n_1 = -\infty}^{\infty} \sum_{k=1}^{K} 
p_k(\mathbf{n}+\boldsymbol{\beta})\sum_{j_k=1}^{J_k}
\delta(\mathbf{r-r_n(\boldsymbol{\alpha},{\boldsymbol\beta})-r}_{j_k}),
\end{align}
\begin{align}
\label{math/29}
p_k(\mathbf{n}+\boldsymbol{\beta})&=\left| \begin{array}{rr}
1,& \; \:\boldsymbol{\Upsilon}_{\mathbf n}(\boldsymbol{\beta})\in{U_k} , \\
0,& \;\boldsymbol{\Upsilon}_{\mathbf n}(\boldsymbol{\beta}) \notin{U_k} ,  
\end{array} \right.  
\end{align}
where ${\mathbf{r}_j}_{k}$ is a 
vector, determining the location of the atom centres of the structural cell 
with the index $k$ relative to the nodes of the lattice, to which the cell 
is related, the areas $U_k$ are placed in a rectangle 
$\sigma_0 \times \sigma_1$. In general case the area $U_k$ 
will be an arbitrary form (fig.\,2).
\begin{figure}[h]
\centering             
\includegraphics*{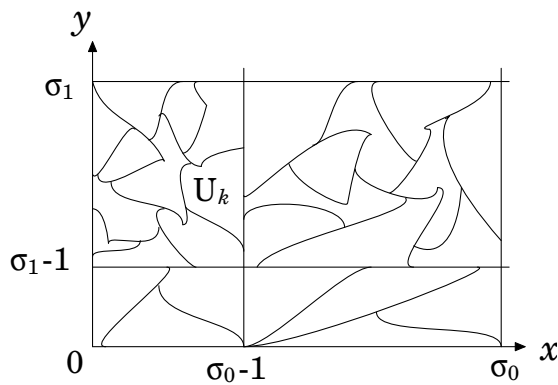}
\\ \vspace{3pt} 
\caption{In the area of U$_k$, corresponding structural 
cell with the index $k$, the condition 
\mbox{$p_k(\mathbf{n}+\boldsymbol{\beta})=1$} is true.}
\label{f:2}
\end{figure}
\\ 
The area U$_k$ can be defined if to compute  
{$\sim$ 100 $\div$ 200} points 
${\boldsymbol\Upsilon}_{\mathbf n}(\boldsymbol{\beta})$ for each 
structural cell. 
Equations (3) and (4) define the atom locations of the quasicrystal and  
can be easily inserted into routine{\footnote{The computer program-generator 
of the quasicrystal structures of Penrose and Ammann-Beenker can be downloaded
http://www.kirensky.ru/download/tilings.zip}} to generate  
quasicrystal structures of any size with arbitrary parameters 
$\boldsymbol{\alpha}$, $\boldsymbol{\beta}$.
As an example we'll present the solution for quasicrystals of Penrose [17] 
($\sigma_0\!=\!\sigma_1\!=\!\chi_0\!=\!\chi_1\!=\!\tau$, 
$\tau\!=\!(1+\sqrt{5})/2$, $a_0\!=\!a_1\!=\!1$) and 
Ammann-Beenker [18, 19]
\mbox{($\sigma_0\!=\!\sigma_1\!=\!\chi_0\!=\!\chi_1\!=a_0\!=\!a_1\!=\!\sigma$, 
$\sigma\!=\!1+\sqrt{2}$ )} (fig.3). The analytical formulas of areas $U_k$ 
given in appendix.
On the fig.\,4 superposition of two Penrose quasicrystals
with different parameters $\boldsymbol{\beta}$ is presented.
The routine by formulas (3) and (4) is inserted directly into graphical 
file generating fig.\,4 and parameters $\boldsymbol{\beta}$ change randomly for
each loading the figure so one can observe phason [20-23] flips in real time 
zooming screen or loading several windows.
\begin{figure}[pt]
\centering             
\includegraphics{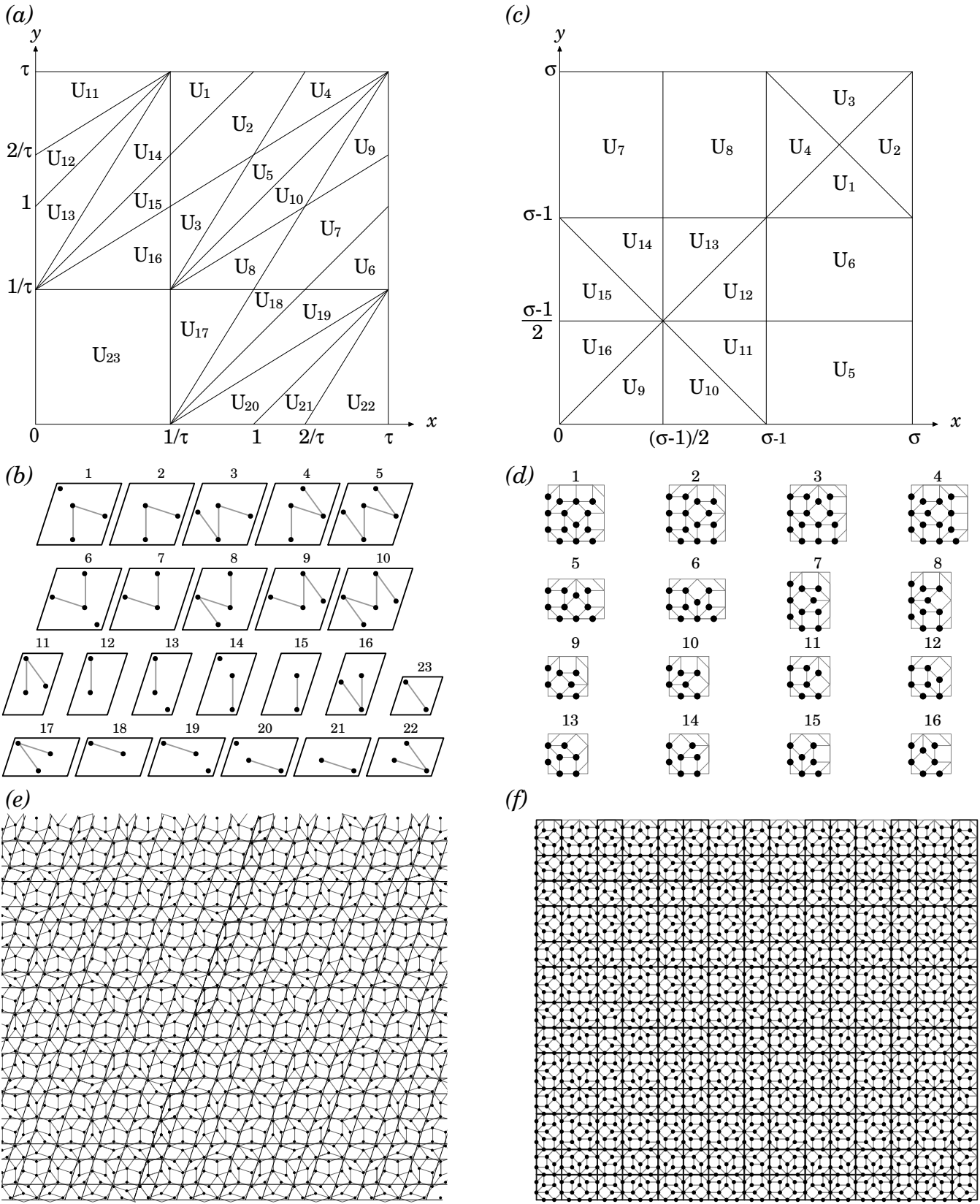}
\caption{In the upper left part of the figure $(a)$ the card of the areas U$_k$ and 
corresponding $(b)$ 23 structural cells for the Penrose quasicrystal are placed, 
in the right upper part -- a similar scheme $(c)$, $(d)$ for the Ammann-Beenker 
quasicrystal with sixteen structural cells. 
The quasicrystals $(e)$, $(f)$ in the lower part of the figure are generated by 
the computer program based on formulas (3), (4) and schemes $(a)$, $(b)$
and $(c)$, $(d)$ correspondingly. The analytical formulas of areas $U_k$ 
for the cards $(a)$, $(c)$ given in appendix.}
\label{f:3}
\end{figure}
$ $
\begin{figure}[pt]
\centering             
\includegraphics*{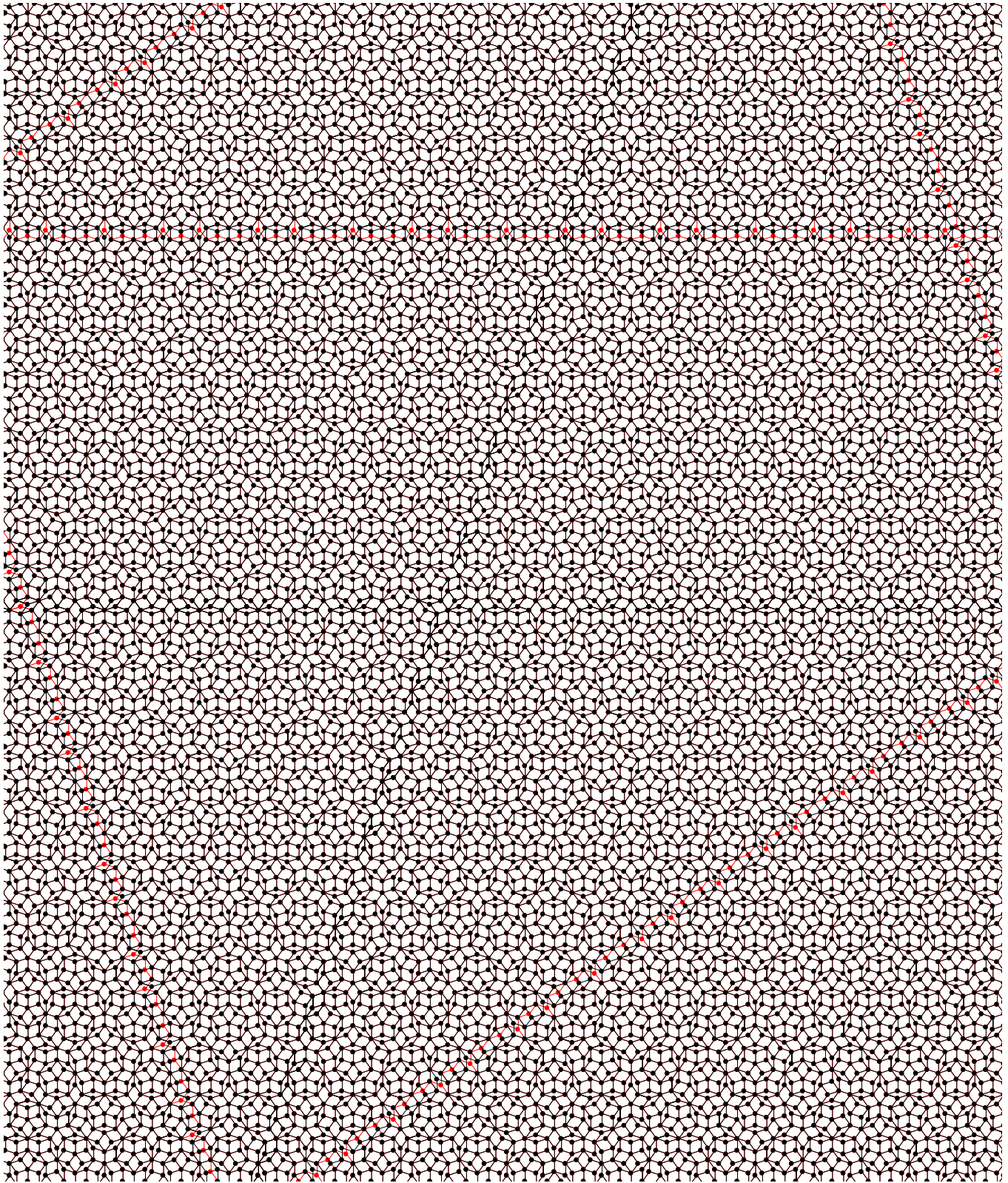}
\caption{
Phason flips in real time.
The routine by formulas (3) and (4) is inserted directly into graphical 
file generating fig.\,4 and parameters $\boldsymbol{\beta}$ change randomly for
each loading the figure so one can observe phason flips zooming screen or 
loading several windows. (This file must be $\ast$.ps not $\ast$.pdf)}  
\end{figure}
\\ \indent The method of describing quasicrystal structures was for demonstration 
presented on the plane. 
In the three--dimensional space the structural cells will be parallelepipeds 
with ribs, which are
parallel with the crystallographic axes, the areas U$_k$ will be the elements
of volume in a rectangular parallelepiped $\sigma_0\times\sigma_1\times\sigma_2$.
For the first time this approach is proposed in [24], the exact solution for the 
quasicrystal of Penrose given in [25].
\section       {\normalsize Rotational symmetries}
%
Analytical calculation [25] of diffraction radiating by formula (3) and (4) 
for the arbitrary form of the areas U$_k$ shows  
that the diffraction patterns do 
not depend on the parameters $\boldsymbol{\alpha}$ and $\boldsymbol{\beta}$.
Since the physical property of a quasicrystal, radiation scattering, is 
independent on $\boldsymbol\alpha$ and $\boldsymbol\beta$, 
the condition of invariance for quasicrystal will be the following [15, 16]:
if under the operator influence of the transformation 
on the atom density distribution 
$\rho(\mathbf{r},\boldsymbol{\alpha},\boldsymbol{\beta})$ only the 
$\boldsymbol{\alpha}$ and $\boldsymbol{\beta}$ parameters change, but not
the $\rho$ function itself
then the quasicrystal is invariant in respect to such a transformation.
Thus, if for the quasicrystal the following condition holds true 
\vspace{-1pt}
\begin{equation}
\label{math/6}
R(2\pi/q,\mathbf{r}_0)\rho(\mathbf{r},\boldsymbol{\alpha},\boldsymbol{\beta})=
\rho(\mathbf{r},\boldsymbol{\alpha}^{\,\prime},\boldsymbol{\beta}^{\,\prime}),
\end{equation}
where $R(\varphi,\mathbf{r}_0)$ -- the rotation operator, $\varphi$ -- 
the corner of rotation, $\mathbf{r}_0$ -- is the point coordinate, 
around which the rotation is performed, then such a quasicrystal will have 
the symmetry axis of the $q$ order (Fig.\,5).
\begin{figure}[h]
\centering             
\includegraphics*{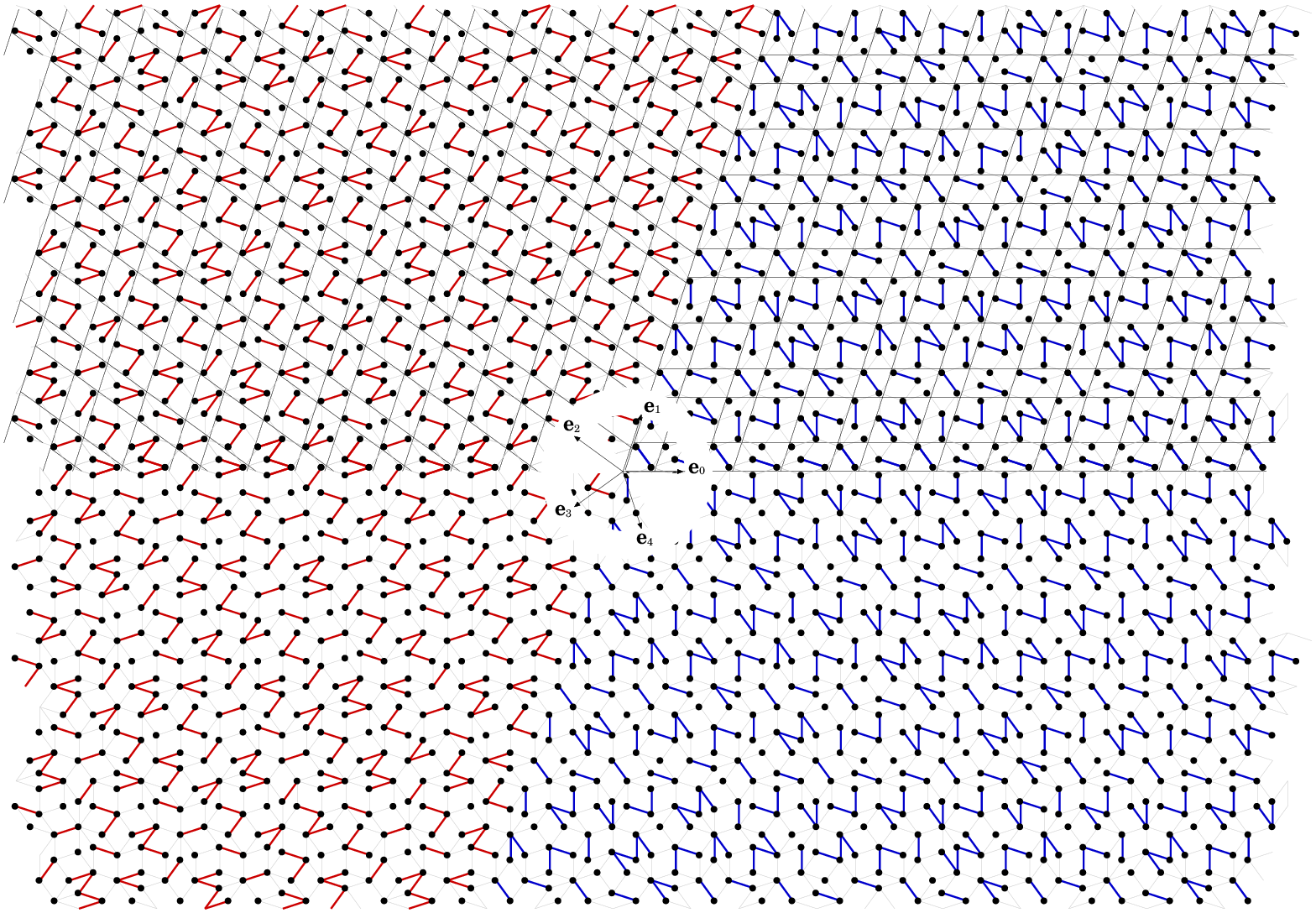}
\caption{Illustration of symmetry on the example of the Penrose quasicrystal. 
In the coordinate bases 
$\mathbf{e}_1,\mathbf{e}_2$ and $\mathbf{e}_0,\mathbf{e}_1$ 
the quasicrystal will be described by formulas (3), (4) but with different 
parameters \mbox{$\boldsymbol{\alpha}$ and $\boldsymbol{\beta}$.}}
\vspace{-10pt}
\end{figure}
%
\section                {\normalsize Conclusion}
\vspace{-7pt}
To describe quasicrystals, in contrast to periodic crystals, use is made of a
concept on qyasicrystals as objects consisting of several structural units 
with ribs, which are parallel with crystallographic axes and
having a quasiperiodic ordering. It is due to the quasiperiodic ordering of 
structural units that symmetries, different from classical, are permissible for
quasicrystals. Such approach gives simple and explicit analytical formulas 
which describe the location of each atom in a quasicrystal.
%
%
%
\appendix
\renewcommand{\thesection}{\normalsize Appendix \Alph{section} }

\section      {\normalsize  The analytical formulas of areas $U_k$ 
                for the Penrose quasicrystal}
Designations $x=\tau \{(n_0+\beta_0)/\tau\}$, $y=\tau \{(n_1+\beta_1)/\tau\}$
are used hereinafter.
\begin{multicols}{2}
\begin{subequations}
\allowdisplaybreaks
\begin{align*}
U_1=  
\bigl(&x\geqslant 1/\tau\bigr)\cap\bigl(y < \tau\bigr)
\cap\bigl( y > x+1/\tau\bigr) , \\ \\
U_2=
 \bigl(&x\geqslant 1/\tau\bigr)\cap\bigl(y< \tau\bigr)
\cap\bigl(y < x+1/\tau\bigr) \\
&\cap\bigl(y> x/\tau+1/\tau\bigr)\cap\bigl(y >\tau x-1/\tau^2\bigr) , \\ \\
U_3=
 \bigl(&x\geqslant 1/\tau\bigr)\cap\bigl(y < x/\tau + 1/\tau\bigr)\\
&\cap\bigl(y> \tau x-1/\tau^2\bigr) ,\\ \\
U_4=
 \bigl(&y< \tau\bigr)\cap\bigl(y > x/\tau - 1/\tau\bigr) \\
& \cap\bigl(y < \tau x-1/\tau^2\bigr) ,\\ \\
U_5=
 \bigl(&y> x\bigr)\cap \bigl(y < x/\tau - 1/\tau\bigr) \\
&\cap\bigl(y< \tau x-1/\tau^2\bigr) ,\\ \\
U_6=
\bigl(&x<\tau\bigr)
\cap\bigl(y \geqslant 1/\tau\bigr)\cap\bigl(y < x-1/\tau\bigr) ,\\ \\
U_7=
\bigl(&x< \tau\bigr)  
\cap\bigl(y \geqslant 1/\tau\bigr)\cap\bigl(y> x-1/\tau\bigr) \\
&\cap\bigl(y< x/\tau + 1/\tau^3\bigr)\cap\bigl(y> \tau x-1\bigr) ,\\ \\
U_8=
\bigl(&y\geqslant 1/\tau\bigr)\cap\bigl(y< x/\tau+1/\tau^3\bigr)\\
&\cap\bigl(y> \tau x-1\bigr) ,\\ \\
U_9=
\bigl(&x < \tau)\cap\bigl(y > x/\tau+1/\tau^3\bigr) \\
&\cap\bigl(y< \tau x-1\bigr) ,\\ \\
U_{10}\!=&
\bigl(x\!>\!y\bigr)\!
\cap\!\bigl(y\!>\!x/\tau\!+\!1/\tau^3\bigr)\!\cap\!\bigl(y>\!\tau x\!-\!1\bigr) ,\\ \\ 
U_{11}=&
\bigl(x \geqslant 0)  \cap ( y<\tau\bigr)\cap\bigl(y > x/\tau+2/\tau\bigr),\\ \\
U_{12}=&  
 (x\!\geqslant\!0\bigr)\!\cap\!\bigl(y\!<\!x/\tau\!+\!2/\tau\bigr)
 \!\cap\!\bigl(y\!>\!x\!+\!1\bigr) ,\\ \\
U_{13}=&
 (x \geqslant 0)\cap\bigl(y > \tau x + 1/\tau\bigr) 
 \cap\bigl(y < x+1\bigr), \\ \\
U_{14}=&
\bigl(x\!<\!\tau\bigr)\!\cap\!\bigl(y\!<\!\tau x\!+\!1/\tau\bigr)\!
\cap\!\bigl(y\!>\!x\!+\!1/\tau\bigr) ,\\ \\
U_{15}=&
\bigl(x\!\geqslant\!\tau\bigr)\!\cap\!\bigl(y\!<\!x\!+\!1/\tau\bigr)  
\!\cap\!\bigl(y\!>\!x/\tau\!+\!1/\tau\bigr) ,\\ \\
U_{16}=&
\bigl(x\!\geqslant\!1/\tau\bigr)\!\cap\!\bigl(y\!\geqslant\!1/\tau\bigr) 
\!\cap\!\bigl(y\!<\!x/\tau\!+\!1/\tau\bigr) ,\\ \\
U_{17}=&
\bigl(y < 1/\tau\bigr)\cap\bigl(x \geqslant \tau\bigr) 
\cap\bigl(y > \tau x-1\bigr) , \\ \\
U_{18}=&
\bigl(y\!<\!1/\tau\bigr)  
\!\cap\!\bigl(y\!>\!x\!-\!1/\tau\bigr)\!\cap\!\bigl(y\!<\!\tau x\!-\!1\bigr) ,\\ \\
U_{19}=&
\bigl(y\!<\!1/\tau\bigr)\!\cap\!\bigl(y\!<\!x\!-\!1/\tau\bigr) 
\!\cap\!\bigl(y\!<\!x/\tau\!-\!1/\tau^2\bigr) ,\\ \\
U_{20}=&
\bigl(y\!\geqslant\!0\bigr)\cap\bigl(y\!>\!x\!-\!1\bigr) 
\cap\bigl(y\!<\!x/\tau\!-\!1/\tau^2\bigr) ,\\ \\
U_{21}=&
\bigl(y \geqslant 0\bigr)\cap\bigl(y < x -1\bigr) 
\cap\bigl(y > \tau x-2\bigr) ,\\ \\
U_{22}=&
\bigl(y \geqslant 0\bigr)\cap\bigl(x<\tau\bigr) 
\cap\bigl(y < \tau x-2\bigr) ,\\ \\
U_{23}=&
\bigl(0\leqslant x < 1/\tau\bigr)\cap\bigl(0 \leqslant y < 1/\tau\bigr) .
\end{align*}
\end{subequations}
\end{multicols}
\pagebreak
\section      {\normalsize The analytical formulas of areas $U_k$ for the  \\
               Ammann-Beenker quasicrystal}
Designations $x=\sigma \{(n_0+\beta_0)/\sigma\}$, $y=\sigma \{(n_1+\beta_1)/\sigma\}$
are used hereinafter.
\begin{multicols}{2}
\begin{subequations}
\allowdisplaybreaks
\begin{align*}
U_1=&
\bigl(y\!>\!\sigma\!-\!1\bigr)\!\cap\!\bigl(y\!<\!x\bigr)
     \!\cap\!\bigl(y\!<\!2\sigma\!-\!1\!-\!x\bigr) ,\\ \\
U_2=&
\bigl(y\!>\!\sigma\!-\!1\bigr)\!\cap\!\bigl(y\!<\!x\bigr)
     \!\cap\!\bigl(y\!>\!2\sigma\!-\!1\!-\!x\bigr) ,\\ \\
U_3=&
\bigl(y\!>\!\sigma\!-\!1\bigr)\!\cap\!\bigl(y\!>\!x\bigr)
     \!\cap\!\bigl(y\!>\!2\sigma\!-\!1\!-\!x\bigr) ,\\ \\
U_4=&
\bigl(y\!>\!\sigma\!-\!1\bigr)\!\cap\!\bigl(y\!>\!x\bigr)
     \!\cap\!\bigl(y\!<\!2\sigma\!-\!1\!-\!x\bigr) ,\\ \\
U_5=&
\bigl(y\leqslant(\sigma-1)/2\bigr)\cap\bigl(x>\sigma-1\bigr),\\ \\
U_6=&
\bigl(y>(\sigma-1)/2\bigr)\cap\bigl(x>\sigma-1\bigr),\\ \\
U_7=&
\bigl(x\leqslant(\sigma-1)/2\bigr)\cap\bigl(y>\sigma-1\bigr),\\ \\
U_8=&
\bigl(x>(\sigma-1)/2\bigr)\cap\bigl(y>\sigma-1\bigr) ,\\ \\
U_9=&
\bigl(x>y\bigr)\cap\bigl(x\leqslant(\sigma-1)/2\bigr) ,\\ \\
U_{10}=&
\bigl(x>(\sigma-1)/2\bigr)\cap\bigl(y<\sigma-1-x\bigr) ,\\ \\
U_{11}=&
\bigl(x\!\leqslant\!\sigma\!-\!1\bigr)\!\cap\!\bigl(y\!>\!\sigma\!-\!1\!-\!x\bigr)
        \!\cap\!\bigl(y\!\leqslant\!(\sigma\!-\!1)/2\bigr) ,\\ \\
U_{12}=&
\bigl(x\leqslant\sigma-1\bigr)\cap\bigl(y<x\bigr)\cap\bigl(y>(\sigma-1)/2\bigr) ,\\ \\
U_{13}=&
\bigl(y\leqslant\sigma-1\bigr)\cap\bigl(y>x\bigr)\cap\bigl(x>(\sigma-1)/2\bigr) ,\\ \\
U_{14}=&
\bigl(y\!\leqslant\!\sigma\!-\!1\bigr)\!\cap\!\bigl(y\!>\!\sigma\!-\!1\!-\!x\bigr)
\!\cap\!\bigl(x\!\leqslant\!(\sigma\!-\!1)/2\bigr) ,\\ \\
U_{15}=&
\bigl(y>(\sigma-1)/2\bigr)\cap\bigl(y<\sigma-1-x\bigr) ,\\ \\
U_{16}=&
\bigl(y>x\bigr)\cap\bigl(y\leqslant(\sigma-1)/2\bigr) .
\end{align*}
\end{subequations}
\end{multicols}
\noindent It is necessary to replace "$\cap$" by "and" to insert analytical 
formulas from the appendix A and B into routine to generate the quasicrystals.

\renewcommand{\refname}{}


{\normalsize References}
\begin{thebibliography}{}
\small
\bibitem{1} D. Shechtman, I. Blech, D. Gratias, J.W. Cahn: Phys. Rev. Lett. 
            1984. V. 53. P. 1951.
\vspace{-6pt}
\bibitem{2} D. Gratias: Les quasi-cristaux//La Recherche. 178. (Juin 1986) 788.
\vspace{-6pt}
\bibitem{3} P. Kramer: Acta Crystallogr., 1982. V. A38. P. 257-264.
\vspace{-6pt}
\bibitem{4} P. Kramer, R. Neri: Acta Crystallogr., 1984. V. A40, P. 580.
\vspace{-6pt}
\bibitem{5} P. A. Kalugin, A. Kitaev, and L. Levitov: JETP 1985. V. 41, P. 119.
\vspace{-6pt}
\bibitem{6} M.Duneau and A. Katz: Phys. Rev. Lett., 1985. V. 54, P. 2688.
\vspace{-6pt}
\bibitem{7} T. Janssen: Europhys. Lett. 1996. V. 14. P. 131. 
\vspace{-6pt}
\bibitem{8} N. de Bruijn:  Ned. Akad. Weten. Proc. Ser. A, 1981. V. 43. P. 27; 
           1981. V. 43. P. 39; 1981. V. 43. P. 53.\\
\vspace{-18pt}
\bibitem{9} J.E.S. Socolar, P. J. Steinhardt, and D. Levine: Phys. Rev. B. 1985. 
          V. 32. P. 5547.
\vspace{-6pt}
\bibitem{10} D. Levine., P. J. Steinhardt: Phys. Rev. B, 1986. V. 34. P. 596.
\vspace{-6pt}
\bibitem{11} F. Gahler and J. Rhyner: Math. Phys. A. 1986. V. 19. P. 267.
\vspace{-6pt}
\bibitem{12} H. -C. Jeong \& P.J. Steinhardt:  Phys. Rev. Lett., 1994. V. 73. PP. 
           1943-1946.
\vspace{-6pt}
\bibitem{13} J. X. Zhong and R. Mosseri: J. Phys. C, 1995. V. 7. P. 8383. 
\vspace{-6pt}
\bibitem{14} P. Repetowicz, U. Grimm, and M. Schreiber: Phys. Rev. B, 
             1998. V. 58. P. 13482. 
\vspace{-6pt}
\bibitem{15} A. Janner, T. Janssen: Phys. Rev. B. 1977. V. 15. P. 643. 
\vspace{-6pt}
\bibitem{16} A. Janner, T. Janssen: Acta Crystallogr. A.  1980. V. 36. P. 399.
\vspace{-6pt}
\bibitem{17} R. Penrose: Bull. Inst. Math. Appl., 1974. V. 10. P. 226.
\vspace{-6pt}
\bibitem{18} P. M. Beenker: Univ. of Technology, Eindhoven T. H., Report WSK 1982.
\vspace{-6pt}
\bibitem{19} R. Ammann, B. Grunbaum, G. C. Shephard: 1992. Discrete Comput. Geom. 
8, 1 
\vspace{-6pt}
\bibitem{20} L.-H. Tang: 
\newblock Phys. Rev. Lett. 1990. V. 64. P. 2390.
\vspace{-6pt}
\bibitem{21}
L.~J. Shaw, V.~Elser, and C.~L. Henley:
\newblock Phys. Rev. B. 1991. V. 43. P. 3423.
\vspace{-6pt}
\bibitem{22}
M.~E.~J. Newman and C.~L. Henley:
\newblock Phys. Rev. B. 1995. V. 52. P. 6386.
\vspace{-6pt}
\bibitem{23}
M.~de~Boissieu {\em et~al.}:
\newblock Phys. Rev. Lett. 1995. V. 75. P. 89.
\vspace{-6pt}
\bibitem{24} V.C. Gouliaev : Quasicrystals. (Part 1)//Russia. Krasnoyarsk.
      Krasnoyarsk State Pedagogical  \\
$\phantom{*}$ University. Preprint 1F (1994)  22 p.
\vspace{-6pt}
\bibitem{25} V. Gulyaev (Gouliaev) : e-print cond-mat/9903263 (1999).
\end{thebibliography}
\end{document}